\documentclass{cjaa}                    
\usepackage{graphicx}                   
\newcommand{\Msolar}{\mbox{\,$\rm M_{\odot}$}}        

\begin{document}
\title{The Relation between the Inclinations of Broad Line
Regions and the Accretion Disk}
\author{Wei-Hao Bian}
\institute{$^{1}$Department of Physics, Nanjing Normal University,
Nanjing 210097, China
\\
$^{2}$National
Astronomical Observatories, Chinese Academy of Sciences, Beijing
100012, China}

\date{Received ...; accepted ...}
\titlerunning{}
\authorrunning{W. Bian}

\abstract{According to the standard model, an active galactic
nucleus (AGN) consists of an inner accretion disk with a jet
around a central massive black hole, and a number of outer broad
line regions (BLRs) and narrow line regions (NLRs). The
geometrical relationship between the BLRs and the accretion disk
is not well understood. Assuming the motion of the BLRs is
virialized and its configuration is disk-like, we derived its
inclination to the line of sight for a sample of AGNs from their
bulge stellar velocity dispersion, their size of the BLRs and
their H$\beta$ linewidth. Compared with the inclination of the
accretion disk obtained from the X-ray Fe K$\alpha$ emission
lines, we found that there is no positive correlation between the
two. Our results showed that BLRs are not coplanar with the
accretion disk and that we should be cautious of using the BLRs
inclination as the disk inclination. The non-coplanar geometry of
the outer BLRs and the inner accretion disk provides clues to the
origin of BLRs and the properties of the accretion disk. Our
preferable interpretation is that BLRs arise out of the outer part
of a warped accretion disk. 
          }
\maketitle

\section{Introduction}
One basic component of the standard model of active galactic
nuclei (AGNs) is an inner accretion disk around a supermassive
black hole. Then there are broad-line regions (BLRs) and
narrow-line regions (NLRs) outside the accretion disk, which are
responsible for the emission lines appearing in the AGN spectra.
The geometrical relation between the different components is
pertinent to an understanding of the physics of the AGNs. Many
lines of evidence show that there is little or no correlation
between the position of the radio jets and the major axis of the
disk of the host galaxy (Schmitt et al. 2001). However, for the
inner region of AGNs the geometrical relation between the BLRs and
the accretion disk still remains a mystery. The reason is that it
is difficult to determine the inclinations of the BLRs and the
disk.

Recently Wu \& Han (2001) suggested a simple method to calculate
the BLR inclinations in AGNs with the virial reverberation masses
($M_{\rm rm}$) and the bulge stellar velocity dispersion
($\sigma$). Up to now, there are just 37 AGNs with measured
$M_{\rm rm}$  (Ho 1998; Wandel et al. 1999; Kaspi et al. 2000) and
about a dozen AGNs with measured $\sigma$ (Wu \& Han 2001, and
reference therein). Fortunately there exists an empirical
size-luminosity relation (Kaspi et al. 2000), which can be used to
calculate the virial mass. At the same time, we can estimate
$\sigma$ from [O~III] linewidth because there exists a strong
correlation between $\sigma$ and [O~III] linewidth (Nelson \&
Whittle 1996).

It is difficult to determine the accretion disk inclinations in
AGNs. Comparing the theoretical spectra from the standard
accretion disk with the observed optical spectra, Laor (1990)
found his derived accretion disk inclinations have large
uncertainties. Using X-ray Fe K$\alpha$ profiles, Nandra et al.
(1997) derived the inclinations of the accretion disk for a sample
of 18 Seyfert galaxies observed by ASCA.

In this paper we use the sample of Nandra et al. (1997) to
investigate the relation between the inclinations of BLRs and the
accretion disk. In the next section we describe the data and the
methods to derive these two inclinations. Our result and
discussion are presented in section 3. All of the cosmological
calculations in this paper assume $H_{0}=75~ \rm km~ s^
{-1}~Mpc^{-1}$, $\Omega =1.0$, $\Lambda=0$.

\section{Data}
\subsection{Disk Inclinations}
X-ray emission of many AGNs consists of a power law component, a
soft X-ray excess at lower X-ray energies, a strong Fe emission
line at about 6.4 keV, and the compton-reflection hump in the
energy range of 20-100 keV (Fabian 2000). The broad Fe K$\alpha$
line at 6.4 keV is believed to arise from the fluorescence of the
neutral iron in the inner regions of the accretion disk (Reynolds
\& Fabian 1997), which could give the information about disk
inclinations. Nandra et al. (1997) presented a sample of 18
Seyfert galaxies and fitted the broad Fe K$\alpha$ line observed
by ASCA using the models proposed by Fabian et al. (1989) and Laor
(1991). The inclinations of the accretion disk for these 18 AGNs
(Nandra et al. 1997) are listed in table 1. We adopted the
inclinations from three models: Model A, the Schwarzschild mode;
Model B, the Schwarzschild mode with q=2.5, where q is a parameter
about the line emissivity as a function of radius; Model C, the
Kerr model (Nandra et al. 1997).

\subsection{BLRs Inclination}
Here we use the method proposed by Wu \& Han (2001) to obtain the
inclinations of BLRs. If BLRs is disk-like, the FWHM of H$\beta$
($V_{\rm FWHM}$) is given by (Wills \& Browne 1986)
\begin{equation}
V_{\rm FWHM}=2(V_{\rm r}^{2}+V_{\rm p}^{2}sin^{2}\theta)^{1/2},
\end{equation}
where $\theta$ is the BLRs inclination, $V_{\rm p}$ is the
component in the plane of the disk, and $V_{\rm r}$ is the random
isotropic component. Because $V_{r}$ is usually believed to be
smaller than $V_{p}$, we ignored $V_{r}$ in our calculation (Zhang
\& Wu 2002). If we assume the BLRs motion around the central black
hole is Keplerian, the central black hole mass is
\begin{equation}
M_{\rm bh}=R_{\rm BLR} V_{\rm p}^{2} G^{-1},
\end{equation}
where $G$ is the gravitational constant, and $R_{\rm BLR}$ is the
size of BLRs. The black hole mass can also be derived from the
velocity dispersion (Tremaine et al. 2002),
\begin{equation}
M_{\rm bh}=10^{8.13}(\frac{\sigma}{200 {\rm km~
s^{-1}}})^{4.02}\Msolar, \label{Msigma}
\end{equation}
where $\sigma$ is the bulge stellar velocity dispersion.

From above equations, we can calculate BLRs inclinations knowing
$V_{\rm FWHM}$, $\sigma$, and $R_{\rm BLR}$,
\begin{equation}
sin\theta=\sqrt{\frac{V_{1000}^2R_{\rm
day}}{2768.2\sigma_{200}^{4.02}}},
\end{equation}
where $V_{1000}=V_{\rm FWHM}/(1000 \rm kms^{-1})$, $R_{\rm day}$
is the BLRs size in units of light day, and $\sigma_{200}$ is the
stellar velocity dispersion in units of 200 $\rm km~s^{-1}$.

The BLRs sizes can be calculated from the reverberation mapping
method or from the size-luminosity empirical formula (Kaspi et al
2000),
\begin{equation}
R_{\rm day}=32.9(\frac{\lambda L_{\lambda}(5100~
\rm{\AA})}{10^{44} \rm erg~ s^{-1}})^{0.7} ~~\rm{lt-d},
\end{equation}
where $\lambda L_{\lambda}(5100\rm{\AA)}$ can be estimated from
the $B$ magnitude (Veron-Cetty \& Veron 2001b) by adopting an
average optical spectral index of -0.5 and accounting for the
galactic redding and $k$-correction (Wang \&  Lu  2001; Bian \&
Zhao 2003a,2003b,2004). The stellar velocity dispersions are
adopted from Wu \& Han (2001) and the references therein. If the
stellar velocity dispersion is not available in the literature, we
derived it from [O~III] linewidth. The calculated inclinations of
BLRs for the sample of Nandra et al. (1997) are listed in table 1.

\begin{figure}
\begin{center}
\centerline{\includegraphics{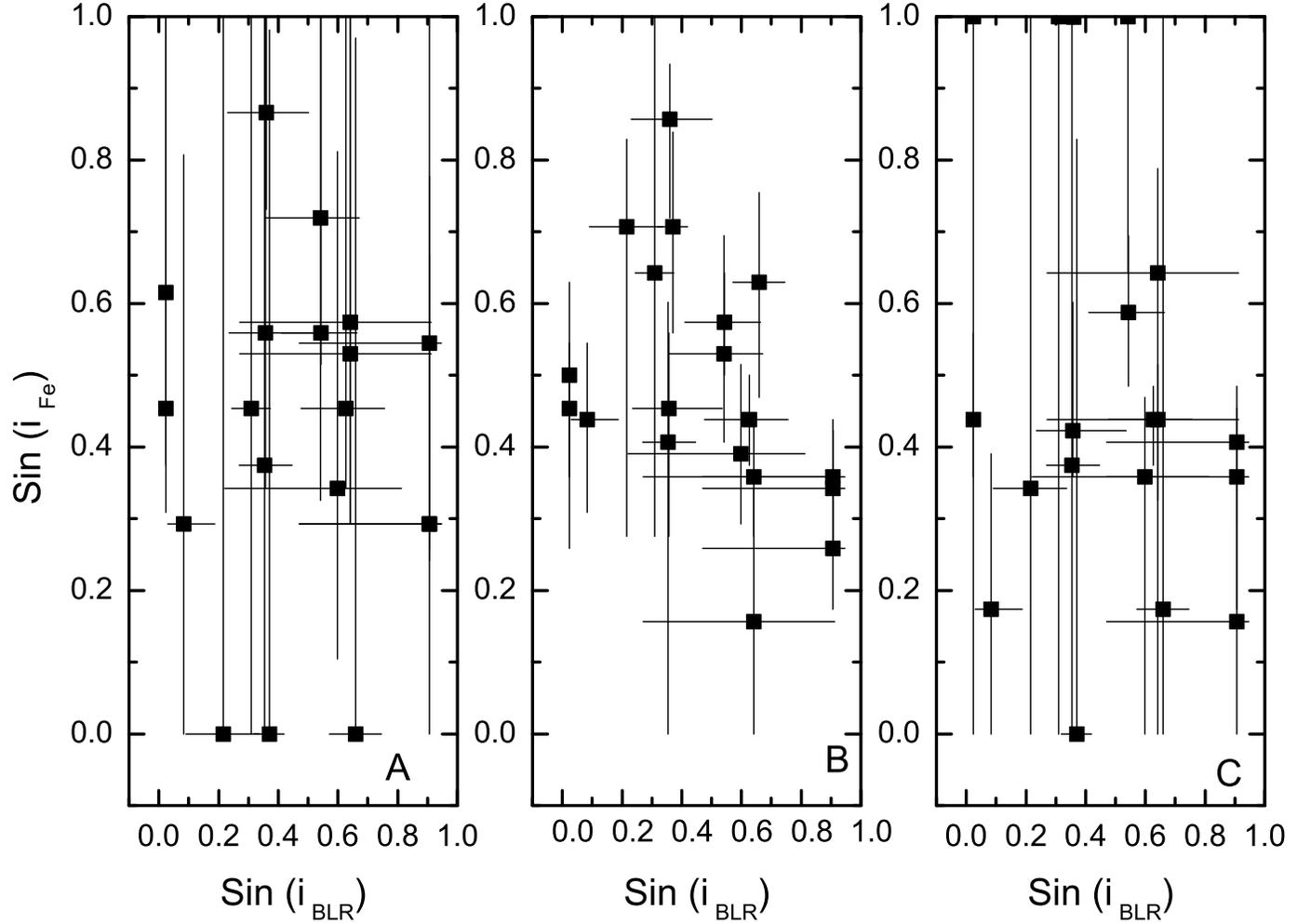}}
\caption{The inclination of
the accretion disk versus the inclination of BLRs. }
\end{center}
\end{figure}

\begin{table*}
\caption{Inclinations of BLRs and the accretion disk}
\begin{center}
\begin{tabular}{lllllllllll}
\hline\hline

Name & $i_{Fe}$(A) & $i_{Fe}$(B) & $i_{Fe}$(C) & $R_{BLR}$ & FWHM
& $\sigma$ & $i_{BLR}$ \\
(1)&(2)&(3)&(4)&(5)&(6)&(7)&(8)\\
\hline
Mrk335        & $  22^{+68}_{-22} $&$ 24^{+13}_{-24} $&$ 22^{+68}_{-22} $&  16.4          & 1620 & 119$^a$   &   $21^{+6}_{-5}$   \\
F9            & $  46^{+44}_{-27} $&$ 32^{+12}_{- 8} $&$ 89^{+ 1}_{-49} $&  16.3          & 5780 & 181$^a$   &   $33^{+9}_{-12}$   \\
3C120         & $  60^{+30}_{-13} $&$ 59^{+10}_{-13} $&$ 88^{+ 2}_{- 1} $&  42            & 1910 & 162             &   $21^{+9}_{-8}$   \\
NGC3227       & $  20^{+ 8}_{-14} $&$ 23^{+ 8}_{- 6} $&$ 21^{+ 7}_{-21} $&  10.9          & 4920 & 144             &   $37^{+18}_{-24}$   \\
NGC3516       & $  27^{+ 6}_{- 7} $&$ 26^{+ 4}_{- 4} $&$ 26^{+ 3}_{- 4} $&  7             & 4760 & 124             &   $39^{+10}_{-10}$   \\
NGC3783(1)    & $  35^{+ 2}_{-18} $&$ 21^{+ 5}_{- 5} $&$ 26^{+ 5}_{- 7} $&  4.5           & 3790 & 98$^a$    &   $40^{+26}_{-24}$   \\
NGC3783(2)    & $  32^{+ 3}_{-15} $&$  9^{+11}_{- 9} $&$ 40^{+12}_{-40} $&  4.5           & 3790 & 98$^a$    &   $40^{+26}_{-24}$   \\
NGC4051       & $  34^{+ 3}_{-14} $&$ 27^{+ 7}_{-11} $&$ 25^{+12}_{- 4} $&  6.5           & 1170 & 80              &   $21^{+11}_{-7}$   \\
NGC4151(2)    & $  17^{+12}_{-17} $&$ 20^{+ 5}_{- 5} $&$  9^{+18}_{- 9} $&  3             & 5910 & 93              &   $65^{+44}_{-37}$   \\
NGC4151(4)    & $  33^{+ 2}_{-18} $&$ 21^{+ 5}_{- 6} $&$ 24^{+ 5}_{- 7} $&  3             & 5910 & 93              &   $65^{+44}_{-37}$   \\
NGC4151(5)    & $  17^{+ 5}_{- 3} $&$ 15^{+ 4}_{- 5} $&$ 21^{+ 5}_{-11} $&  3             & 5910 & 93              &   $65^{+44}_{-37}$   \\
Mrk766        & $  34^{+ 3}_{- 3} $&$ 35^{+ 5}_{- 5} $&$ 36^{+ 8}_{- 7} $&  14.8$\dagger$ & 1630 & 94$^b$    &   $33^{+9}_{-9}$   \\
NGC4593       & $  0 ^{+79}_{- 0} $&$ 45^{+12}_{-11} $&$  0^{+56}_{- 0} $&  4             & 3720 & 124             &   $22^{+3}_{-3}$   \\
MCG-6-30-15(1)& $  33^{+ 3}_{- 5} $&$ 34^{+ 5}_{- 5} $&$ 34^{+ 5}_{- 6} $&  3.6$\dagger$  & 1700 & -               &   -    \\
MCG-6-30-15(2)& $  34^{+ 3}_{- 6} $&$ 33^{+ 8}_{-25} $&$ 34^{+16}_{- 9} $&  3.6$\dagger$  & 1700 & -               &   -    \\
IC4329A       & $  17^{+14}_{-17} $&$ 26^{+ 7}_{- 8} $&$ 10^{+13}_{-10} $&  1.4           & 5050 & 234$^a$   &   $5 ^{+6}_{-3}$      \\
NGC5548       & $   0^{+76}_{- 0} $&$ 39^{+10}_{-11} $&$ 10^{+80}_{-10} $&  21.2          & 6300 & 183             &   $41^{+7}_{-6}$      \\
Mrk841(1)     & $  27^{+ 7}_{- 9} $&$ 27^{+ 6}_{- 6} $&$ 26^{+ 8}_{- 5} $&  33.4$\dagger$ & 5470 & 178$^d$               &   $1.4^{+0.2}_{-0.2}$   \\
Mrk841(2)     & $  38^{+ 2}_{-16} $&$ 30^{+ 9}_{-15} $&$ 90^{+ 0}_{-90} $&  33.4$\dagger$ & 5470 & 178$^d$               &   $1.4^{+0.2}_{-0.2}$    \\
Mrk509        & $  27^{+63}_{-27} $&$ 40^{+48}_{-24} $&$ 89^{+ 1}_{-89} $&  76.7          & 2270 & 221$^a$   &   $18^{+4}_{-4}$   \\
NGC7469(2)    & $   0^{+89}_{- 0} $&$ 45^{+11}_{-29} $&$ 20^{+70}_{-20} $&  4.9           & 3000 & 153$^a$   &   $12^{+7}_{-7}$   \\
MCG-2-58-22   & $  46^{+44}_{-46} $&$ 41^{+10}_{-15} $&$ 26^{+64}_{-26} $&  57.8$\dagger$ & 6360 & 155$^c$   &   - \\
\hline
\end{tabular}
\end{center}
Col.1: name, Col.2-4: inclinations of the accretion disk for Model
A, B, and C, respectively, Col.5: the BLRs sizes in units of light
days, Col.6: FWHM of H$\beta$ in units of $\rm kms^{-1}$, Col.7:
the stellar velocity dispersion in units of $\rm
km s^{-1}$, Col.8: inclination of BLRs. \\
$\dagger$: BLRs sizes are calculated from equation 5, the
others are from Kaspi et al. (2000). \\
The velocity dispersions via [O~III] linewidth labelled with $^a$
are from Nelson (2001), labelled with $^b$ are from Veron-Cetty et
al. (2001a), labelled with $^c$ are from Whittle (1992), labelled
with $^d$ are from Wilkes et al. (1999), the others are the
directed measured from host spectra listed in Wu \& Han (2001).
\end{table*}

\begin{table}
\caption{The correlation coefficients and possibility}
\begin{center}
\begin{tabular}{lcccccccccc}
\hline\hline
Model & coefficient & possibility & slop\\
\hline
A & -0.25 & 0.276 & -0.13$\pm$0.4 \\
B & -0.53  & 0.01 & -0.27$\pm$0.1\\
C & -0.25 & 0.28 & -0.24$\pm$0.22 \\
\hline
\end{tabular}
\end{center}
\end{table}

\section{Result and Discussion}
We plot the inclination of the accretion disk versus the
inclination of BLRs. We use the least square linear regression to
fit the data in figure 1, considering the errors of the
inclinations of the accretion disk. The correlation coefficients
and possibility that the correlation is caused by a random factor
are listed in table 2. In Model A and C, there is very weak
correlation between them. In Mode C, there is a median strong
anti-correlation between them, namely, AGNs with a more face-on
accretion disk tend to have larger inclination of BLRs.

Nishiura et al. (1998) used the ratio of the H$\beta$ or H$\alpha$
FWHM to the hard X-ray luminosity to trace the inclination of BLRs
and found there is a negative correlation between inclinations of
BLRs and that of the accretion disk. Here we use the BLRs size
from the reverberation mapping method, the bulge velocity
dispersion, and the FWHM of H$\beta$ to calculate the inclinations
of disk-like BLRs assuming the motion of the BLRs clouds is
virialized (Peterson, Wandel 2000). With the more accurate
estimation of inclination of BLRs, we also find this negative
correlation in Model B, which is consistent with the results of
Nishiura et al. (1998). However, there is very weak correlation in
Model A and C.

We should note that for some objects the BLRs sizes are calculated
from the B-band luminosity or the velocity dispersion is
calculated from the width of [O~III] line. The error of the
calculated inclination of BLRs is mainly from the uncertainties of
size of BLRs, FWHM of H$\beta$, and the velocity dispersion. From
the error transform formula, $\delta \theta=\sqrt{4(\delta
V/V)^2+(\delta R/R)^2+(4.02\delta \sigma /\sigma
)^2}(2tan\theta)$, the error of inclination of 30 (deg) is about
5.5 (deg) assuming 10\% uncertainties of these three parameters
(see Wu \& Han 2001). The uncertainties of BLR sizes and the
H$\beta$ FWHM for NGC 3516 and NGC 4593 are unavailable in the
literature and were assumed to be 10\%. The uncertainties of the
velocity dispersion derived from the [O~III] width were assumed to
be 10\%. The uncertainties of BLRs sizes derived from equation 5
were also assumed to be 10\%. The uncertainties of the calculated
inclinations of BLRs are listed in table 1.

A too large $V_{\rm FWHM}$, a too large $R_{\rm BLR}$, or/and a
too smaller $\sigma$ maybe lead to the right of equation (4)
larger than one, i.e., to the breakdown of the underlying method.
MCG-2-58-22 is a case in point. Using equation (4) to derive the
BLR inclination, we assumed that the BLRs are virialized and the
random isotropic component $v_{r}$ can be omitted. However it is
possible that BLRs are not virialized and the random isotropic
movements can't not be omitted, $(sin\theta)^{2} \propto (V_{\rm
FWHM}^{2}-4V_{\rm r}^{2})R_{\rm BLR}/\sigma^{4.02}$. The
inclination derived from equation (4) is an upper limit when we
omitted the random velocity $v_{r}$. The breakdown of equation (4)
suggested that BLRs are not virialized. The effects of random
velocity on the BLR inclination estimates have been discussed by
zhang \& Wu (2002).

In this paper we adopted the inclinations of the accretion disk
from the fitting of the Fe K$\alpha$ profile with different models
for a sample of 18 Seyfert galaxies observed by ASCA (Nandra et
al. 1997). There is large uncertainties on the inclinations even
for the same model (see table 1.). Recent research showed that the
Fe K$\alpha$ may be some sort of composite feature from inner
accretion disk and/or outer BLRs (Turner et al. 2002). It is
necessary to obtain the Fe K$\alpha$ profile with higher spectral
resolution and have a better model on the origin of Fe K$\alpha$
to derive the inclination of the accretion disk.

In figure 1, we can't find any positive correlation between the
inclination of BLRs and that of the accretion disk, which shows
that BLRs are not coplanar with respect to the accretion disk. The
non-coplanar geometry of outer BLRs and the inner accretion disk
in AGNs provides clues to the property of the accretion disk and
the origin of BLRs (Nicastro 2000; Collin \& Hure 2001; Bian \&
Zhao 2002; Bian \& Zhao 2003c; Laor 2003). There are some ideas to
interpret this non-coplanar geometry in AGNs. Nishiura et al.
(1998) suggested that the BLRs arises from the outer parts of a
warped accretion disk illuminated by the central engine. The
warping of accretion disk can be driven by the radiation pressure.
Recent Nicastro (2000) suggested that the sizes of BLRs are
determined by the transition radius between the radiation pressure
and gas pressure dominated region of the disk. Our calculated
inclinations of BLRs are derived from the disk-like BLRs. If the
BLRs are not disk-like, there is an another suggestion about the
non-coplanar geometry that the gravitational instability of the
standard accretion disk leads to the BLRs (Collin \& Hure 2001;
Bian \& Zhao 2002), which can lead to a sphere-like BLRs
considering the radiation pressure.

Since there is possibly no positive correlation between
inclinations of BLRs and the accretion disk, namely they are not
coplanar, it is a risk to use the inclination of the accretion
disk as the inclination of BLRs. Rokaki \& Boisson (1999)
presented accretion disk fit to the UV continuum and H$\beta$
emission line in a sample of AGNs. They found the inclination
dependence on the central black hole is opposite fitting the UV
continuum and H$\beta$ emission line. They assumed inclinations of
BLRs and that of the accretion disk are the same and then
determined the inclination, black hole mass, and the accretion
rates. We notice that the black hole masses are all larger than
the value of the recent reverberation mapping masses for the
common AGNs (Kaspi et al. 2000). The accretion rates they derived
are also smaller compared to the results of the recent accretion
disk fit to the optical continuum (Collin et al. 2002) and the B
band luminosity (Bian \& Zhao 2002). It is necessary to fit the UV
continuum using the reverberation mapping mass as a known
parameter to constrain other parameters of the accretion disk.

\begin{acknowledgements}
We thank the anonymous referee for valuable comments. This work
has been supported by the NSFC (Nos. 10403005 and 10473005) and
NSF of the Jiangsu Provincial Education Department (No.
03KJB160060).).

\end{acknowledgements}

\end{document}